\documentclass[twocolumn,aps,pra,floatfix]{revtex4-1}
\usepackage{hyperref}
\usepackage{amsmath}
\usepackage{amsfonts}
\usepackage{amsbsy}
\usepackage{xcolor}
\usepackage{soul}
\usepackage{graphicx}

\newcommand{\ket}[1]{\ensuremath{\left\vert #1 \right\rangle}}

\newcommand{\spvec}[1]{\ensuremath{\mathbf{#1}}}

\newcommand{\unitvec}[1]{\hat{\mathbf{{#1}}}}

\newcommand{\cgshort}[3]{\ensuremath{\mathcal{C}_{#1,#2}^{(#3)}}}

\newcommand{\bea}{\begin{eqnarray}}
\newcommand{\eea}{\end{eqnarray}}
\newcommand{\beq}{\begin{equation}}
\newcommand{\eeq}{\end{equation}}

\newcommand{\eq}[1]{(\ref{#1})}

\newcommand{\br}{{\bf r}}

\newcommand{\stochx}{\ensuremath{X}}
\newcommand{\stochxv}{\spvec{\stochx}}

\newcommand{\pola}{\hat{\mathbf{e}}}
\newcommand{\rv}{{\bf r}}

\newcommand{\Dc}{{\cal D}}

\newcommand{\eo}{\epsilon_0}

\renewcommand{\>}{\rangle}

\newcommand{\Pc}{\mathcal{P}}

\newcommand{\commentout}[1]{{}}
\newcommand{\radKernel}{\ensuremath{\mathsf{G}}}

\newcommand{\cbE}{\boldsymbol{\mathbf{\cal E}}}

\begin{document}
\title{Collective resonance fluorescence in small and dense atom clouds:\\ Comparison between theory and experiment}
\date{\today}
\author{S. D. Jenkins}
\affiliation{Mathematical Sciences, University of
  Southampton, Southampton SO17 1BJ, United Kingdom}
\author{J. Ruostekoski}
\affiliation{Mathematical Sciences, University of
  Southampton, Southampton SO17 1BJ, United Kingdom}
\author{J. Javanainen}
\affiliation{Department of Physics, University of Connecticut, Storrs,
Connecticut 06269-3046}
\author{S. Jennewein}
\affiliation{Laboratoire Charles Fabry, Institut d'Optique Graduate School, CNRS, Universit\'e Paris-Saclay, 91127 Palaiseau cedex, France}
\author{R. Bourgain}
\affiliation{Laboratoire Charles Fabry, Institut d'Optique Graduate School, CNRS, Universit\'e Paris-Saclay, 91127 Palaiseau cedex, France}
\author{J. Pellegrino}
\affiliation{Laboratoire Charles Fabry, Institut d'Optique Graduate School, CNRS, Universit\'e Paris-Saclay, 91127 Palaiseau cedex, France}
\author{Y. R. P. Sortais}
\affiliation{Laboratoire Charles Fabry, Institut d'Optique Graduate School, CNRS, Universit\'e Paris-Saclay, 91127 Palaiseau cedex, France}
\author{A. Browaeys}
\affiliation{Laboratoire Charles Fabry, Institut d'Optique Graduate School, CNRS, Universit\'e Paris-Saclay, 91127 Palaiseau cedex, France}

\begin{abstract}
We study the emergence of a collective optical response of a cold and dense $^{87}$Rb atomic cloud to a near-resonant low-intensity light
when the atom number is gradually increased. Experimental observations are compared with microscopic stochastic simulations of recurrent
scattering processes between the atoms that incorporate the atomic multilevel structure and the optical measurement setup. We analyze the optical response
of an inhomogeneously-broadened gas and find that the experimental observations of the resonance line
shifts and the total collected scattered light intensity in cold atom clouds substantially deviate from those of thermal atomic ensembles, indicating strong light-induced
resonant dipole-dipole interactions between the atoms. At high densities, the simulations also predict a significantly slower decay of light-induced excitations
in cold than in thermal atom clouds. The role of dipole-dipole interactions is discussed in terms of resonant coupling examples and the collective radiative excitation eigenmodes
of the system.
\end{abstract}

\maketitle

\section{Introduction}
\label{sec:introduction}

An improved experimental control of cold atoms and an increase in computing power now allow both measurements and many-body simulations of the optical response in small atomic ensembles. These systems
have proved to be important since strong light-induced dipole-dipole (DD) interactions can lead to collective light scattering phenomena.
In particular, it was recently pointed out~\cite{Javanainen2014a} that a cold and dense atomic medium can exhibit light-induced correlations
and an optical response that dramatically differ from those of a thermal, inhomogeneously-broadened medium.

Here we analyze in detail near-resonance light scattering from small clouds of cold or thermal, trapped $^{87}$Rb atoms that were recently experimentally measured
or numerically simulated in Refs.~\cite{Pellegrino2014a,Jenkins_thermshift}. Large-scale numerical classical electrodynamic simulations and
experimental observations indicate the emergence of collective effects in the optical response of a cold-atom sample due to DD interactions,
as the number of atoms is gradually increased and the density of the cloud increases.
The experimentally observed light scattering and classical electrodynamic simulations provide a detailed side-by-side comparison between experiment and theory
in a gas of multilevel $^{87}$Rb atoms.
By performing microscopic
numerical simulations of both cold and hot atomic ensembles,
in which the atoms
are represented by linear point emitters in the low excitation limit,
we find that the experimental observations of the resonance line
shifts and the total collected scattered light intensity substantially
deviate from those of thermal atomic ensembles.
In particular, in both cases the density-dependent resonance shift is absent. However, introducing
inhomogeneous broadening due to thermal atomic motion restores the shift.

The experiments are performed in a microscopic elongated dipole trap where we study the role of DD interactions between $^{87}$Rb atoms in the resonance fluorescence by varying the
atom number from one to $\sim 450$. The sample is illuminated by near-resonant low-intensity laser pulses. Before the illumination
the atoms are laser-cooled to a temperature $\sim 110\mu$K, such that the thermal Doppler broadening is negligible.
The corresponding stochastic simulations go beyond idealized models, by incorporating a nonuniform atom density, the effects of an anisotropic elongated trap,
the multilevel structure of the $F=2$ ground-state manifold of $^{87}$Rb, imaging geometry, and the optical components such as lenses and polarizers.
In the simulations the thermally-induced broadening of hot atoms is generated by stochastically sampling the inhomogeneous
broadening of the resonance frequencies of individual atoms. The numerical simulations also incorporate the recurrent (dependent)
scattering processes between the atoms where the light is scattered more than once by the same atom. These are the source of light-induced correlations
between the atoms~\cite{Javanainen2014a,Ruostekoski1997a}. We find that the strong DD interactions in the system lead to
collective excitation eigenmodes of the system that exhibit a broad range of collective radiative linewidths.
The role of collective eigenmodes is revealed in the temporal profile of the decay of light-induced excitations after the incident laser pulse is switched off.
For large atom numbers in stochastic simulations we find a significantly slower decay of the excitations for cold than for thermal atom clouds.

The radiative interactions in ensembles of resonant emitters constitute an active area of research that covers a variety of systems,
such as  thin thermal atom cells~\cite{Keaveney2012}, metamaterial arrays of nanofabricated resonators~\cite{LemoultPRL10,FedotovEtAlPRL2010,AdamoEtAlPRL2012,CAIT},
arrays of ions~\cite{Meir2013}, and nanoemitters~\cite{Brandes2005,Pierrat2010}. Experiments in cold atomic ensembles have dominantly addressed optically thick but relatively
dilute samples~\cite{Balik2013,Bienaime2010,Loew2005,wilkowski,wilkowski2,Ye2016,Guerin_subr16,vdStraten16,Roof16,Araujo16}, whereas the setup described here
deals with small samples but
at higher densities. It has been used to study fluorescence~\cite{Pellegrino2014a,Jenkins_thermshift} and forward scattering~\cite{Jennewein_trans,antoine_delay}, although the present paper concentrates on the case of fluorescence.
Theoretically, ensembles of resonant emitters provide a rich and challenging phenomenology whereby light-induced correlation effects~\cite{Dicke1954,Lehmberg1970a,Lehmberg1970b,Saunders73,Ishimaru1978,vantiggelen90,Morice1995a,Ruostekoski1997a,
Ruostekoski1997b,Rusek96,Javanainen1999a,Ruostekoski1999a,BerhaneKennedyfer,Clemens2003a,
dalibardexp,JenkinsLongPRB,Jenkins2012a,Olmos13,Castin13,Javanainen2014a,Pellegrino2014a,JavanainenMFT,Kuraptsev14,
Skipetrov14,Bettles_lattice,Lee16,Bettles_prl16,antoine_polaritonic,Longo16,Sutherland16forward,Olmos16,Rey_Weyl,Guerin16b,Rey16} due to
recurrent scattering can profoundly alter the optical response of sufficiently dense samples.

Parallel to our work, the effects of motional dynamics of atoms were observed in a cold Sr atom vapor by comparing the optical
response of narrow and broad linewidth transitions of the atoms~\cite{Ye2016}. A qualitatively different behavior was observed
in the two cases. Unlike our experiment, Ref.~\cite{Ye2016} considered a dilute gas $\rho/k^3\sim 10^{-3}$. Another important difference is that the
narrow Sr resonance is affected by the recoil shift of the atom.

We begin with a brief overview of the experimental setup in Sec.~\ref{sec:experimental-setup}. This is followed by a description of strong
DD interactions in cold and dense atomic ensembles in Sec.~\ref{sec:DD interactions}. We first cover the simple case of a two-level atom and how to
incorporate the effects of inhomogeneous broadening. The discussion is then extended to the multilevel $^{87}$Rb case in Sec.~\ref{sec:87rb-scattering}.
We present the experimental and numerical results in Sec.~\ref{sec:results} by first analyzing the collective modes and their decay  in Sec.~\ref{sec:cooperative-modes},
then the experimental observations in Sec.~\ref{Sec:Exp_results}, and the theoretical results in Sec.~\ref{sec:optic-resp-incid}. Finally, some concluding remarks are
made in Sec.~\ref{sec:conclusion}.

\section{Experimental Setup}
\label{sec:experimental-setup}\label{Sec:Exp_setup}

\begin{figure}
  \centering
  \includegraphics{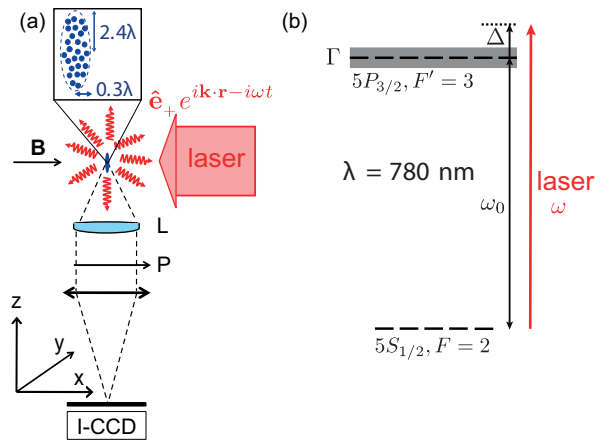}
  \caption{
  (a) Experimental setup. The atoms are
    initially confined in a microscopic single-beam dipole trap (not
    shown) (wavelength $957$ nm, depth $1$ mK, and waist $1.6$ $\mu$m,
    oscillation frequencies $\omega_x = \omega_y = 2\pi \times 62$ kHz
    and $\omega_z=2\pi\times 8$ kHz). The excitation laser propagates
    anti-parallel to  the quantization axis $x$, set by a magnetic
    field $B \sim 1$ G. We collect light
    scattered in the negative $z$ direction, after a polarizer $P$
    oriented at an angle of $55^\circ$ with respect to $x$, using a
    lens $L$ with a large numerical aperture ($\textrm{NA}=0.5$) and
    an image intensifier followed by a CCD camera
    (I-CCD).
    (b) Structure of $^{87}$Rb atoms relevant to this work.
    The excitation light at frequency $\omega$ is near resonant with
    the transition at $\lambda = 2\pi/\omega_0 = 780\,\mathrm{nm}$.
  }
  \label{fig:setup}
\end{figure}

As described in Ref~\cite{Pellegrino2014a},
our experimental setup enables us to access densities and
temperatures at which strong DD interactions
manifest themselves in the optical response.
To achieve this, we laser-cool an ensemble of $N$ $^{87}$Rb atoms in a microscopic
dipole trap obtained by focusing a laser beam with a wavelength of 957 nm onto a spot
size of 1.6~$\mu$m, as depicted in
Fig.~\ref{fig:setup}(a). We obtain an
elongated cloud at a temperature of $\sim
110\,\mathrm{\mu K}$ \cite{Bourgain2013b}, with a thermal Gaussian distribution
with root-mean-square sizes of
$\sigma_x=\sigma_y = 0.3 \lambda$ and $\sigma_z = 2.4\lambda$,
where $\lambda \simeq 780.2\,\mathrm{nm}$ is the resonant wavelength of
the D2 transition used in this work.
Varying the number of atoms in the trap from $N=1$ to $N=450$ allows
us to study the increasing role of inter-atomic interactions
in the radiative response. For the largest atom numbers
the peak density
at the center of the trap is $\rho \simeq 0.9 k^3$ ($k\equiv2\pi/\lambda$), which,
as we show in Sec.~\ref{sec:DD interactions}, results in
DD interactions strong enough to heavily influence the optical response of
the atomic cloud.
At the same time, thermal motion within this cold atomic cloud
produces only a negligible Doppler broadening of $0.04\gamma$, where
$\gamma = 2\pi \times 3\, \mathrm{MHz}$ is the half width half maximum (HWHM) radiative linewidth of the D2 transition
\begin{equation}
  \label{eq:Gamma_E0_def}
  \gamma = \frac{\Dc^2 k^3}{6\pi\hbar\epsilon_0} \, ,
\end{equation}
where ${\cal D}$ denotes the reduced dipole matrix element.
The number of atoms $N$ is experimentally controlled with a 10\% uncertainty. The uncertainties in the temperature, atom number,
and the waist size lead to an uncertainty on the peak atom density of a factor of two.

We observe the light scattered by the cloud when it is illuminated by an incident laser field
tuned near the resonance of the transition between ground and excited hyperfine levels
$|g\>=(5S_{1/2}, F=2)$ and $|e\>=(5P_{3/2},F'=3)$, respectively; Fig.~\ref{fig:setup}(b).
The incident field is sent along the small axis of the cloud and has a $\sigma_+$ polarization,
where we use the unconventional unit vectors,
\begin{equation}
  \pola_{\pm 1} = \mp {1\over \sqrt{2}} (\pola_y \pm i \pola_z), \quad \pola_0 = \pola_x\,,
\label{polarizationvec}
\end{equation}
corresponding
to the labeling of the directions in the experiment, shown in Fig.~\ref{fig:setup}(a), with the $x$ axis  as the quantization axis set by a 1 Gauss magnetic field.

Prior to the illumination by the excitation laser, we prepare
the atoms in the hyperfine ground
level $|g\>$ with $95\%$ efficiency using  a $\sigma_+$-polarized beam tuned on the
$(5S_{1/2}, F=1)$ to $(5P_{3/2},F'=2)$ transition. The atoms are thus prepared
in an incoherent mixture
of Zeeman states $\ket{g,m}$ ($|m|\le2$) with relative
populations $p_m$, estimated by solving rate equations:
\begin{equation}
p_{-2}=p_{-1}\simeq0, \quad p_{0}=p_1=p_2 \simeq 1/3\,.
\label{eq:populations}
\end{equation}

In the experiments reported in Sec.~\ref{Sec:Exp_results} we use flat-top light pulses to excite the prepared atoms, with durations ranging from $T=125\,\mathrm{ns}$ (rise time $\sim 7$ ns, using an electro-optical modulator) to $T=2$~$\mu$s (rise time $\sim 50$ ns, using an acousto-optical modulator).
We employ pulses with low intensity $I<0.1 I_{\rm sat}$
($I_{\rm sat}=1.6\,\mathrm{mW/cm^2})$ in order to operate in the
limit where the response of the atoms to light is linear.
The probe frequency is $\omega
= \omega_0 + \Delta$, where $\omega_0$ is the resonance frequency of
the $|g\>\leftrightarrow |e\>$ transition in the  absence of a magnetic field.

\section{The influence of DD interactions on the optical
  response}
\label{sec:DD interactions}

An incident field nearly resonant with the $ |g\>
\leftrightarrow |e\>$ transition induces electric dipoles in
each of the atoms.  In this Section, we will see how interactions
between the dipoles bring about a collective response of
the gas in response to the incident field. The key here is that each atom is driven not only by the incident field, but also
by the fields emitted by all other atoms in the gas.
The resulting DD interactions profoundly alter the cloud's
scattering properties at atom densities attained in our
experiments.

Generally, in the limit of low light intensity, recurrent scattering, in
which a photon repeatedly scatters between the same atoms, results in
dynamics where a cloud's polarization density depends on two-body
correlations involving the polarization density and the atomic number
density~\cite{Ruostekoski1997a}.
These two-body correlations depend on three-body correlations, which
depend on four body-correlations, and so on.  To exactly account for the
DD interactions in the optical response of an $N$-body
system, one would ultimately need to solve the dynamics of $N$-body
correlation functions.

Fortunately, in the limit of low light intensity one can avoid the coupled equations of the correlation functions by carrying out a stochastic simulation instead~\cite{Javanainen1999a,Lee16}. We first sample stochastically $N$ atomic positions $\{\stochxv_1,\ldots,\stochxv_N\}$ from the joint probability
distribution of the positions.
In our experiment, at temperatures of $110 \, \mathrm{\mu{}K}$, the
de Broglie wavelength is much smaller than the inter-atomic
separation.  We therefore treat the atomic positions in our harmonic
trap as independent identically distributed Gaussian random variables
with the experimentally observed root-mean-square widths $\sigma_x=\sigma_y = 0.3 \lambda$ and $\sigma_z
= 2.4\lambda$. The atom at each position $\stochxv_j$
is taken to be a linear dipole driven by the electric field falling on it.
We next solve the optical response for such a collection of dipoles and compute the relevant observable(s). To model an experiment, real or hypothetical, we finally average the result(s) over
a large number of samples of atomic positions. In the simulations we consider both the steady-state and time-dependent cases.

For each atomic sample, the optical response may be
analyzed in terms of collective dipole excitations, each with a distinct
resonance frequency and decay rate.
The cooperative nature of a cloud's response can be roughly
characterized by the widths of the distributions of collective resonance
shifts and decay rates.
We will show that at the
densities reached in the experiment, the interactions between an atom
and its nearest neighbor sensitively
depend on their relative distance, thus illustrating the role of DD interactions in the formation of
correlations in the cold atoms.

We will then review how this model can be generalized to a cloud of
$^{87}$Rb atoms, accounting for Zeeman degeneracy~\cite{Pellegrino2014a}, in
Sec.~\ref{sec:87rb-scattering}.
In the case of multiple electronic ground states the simulations represent
a classical approximation where the recurrent scattering between the atoms
is included, but nonclassical higher order correlations between internal levels
that involve fluctuations between ground-state coherences of different levels
and the polarization are factorized~\cite{Lee16}.
In the limit of low light intensity (that we assume everywhere in this paper)
the classical approximation
correctly represents all the atomic ground-state population expectation values of
even the exact quantum solution.

We will illustrate that, in a cloud of $N=450$ atoms as realized in the experiment, the
light-induced DD interactions between the atoms result in broad distributions of collective mode resonance frequencies
and decay rates. We will also discuss  the effects of atomic
motion in a thermal sample by introducing an inhomogeneous broadening to the
resonance frequencies of the atoms, but still considering the atoms as frozen during their
interaction with the probe light.

\subsection{Coupled dynamics of two-level atoms}
\label{sec:two-level-atoms}

For simplicity we introduce the theoretical model first by considering an ensemble of two-level atoms, before generalizing to the multilevel scheme that
describes the $^{87}$Rb $F=2$ ground-state experiments.
Since the vectorial nature of electric field and dipole moment matters in this paper, the two-level atom is assumed to have the vector dipole moment matrix
element between the ground and excited state $ \Dc\unitvec{e}$, where $\Dc$ is real and $\unitvec{e}$ is a possibly complex unit vector. This situation could
be approximated experimentally by applying a strong magnetic field on an atom, so that only one transition between the Zeeman levels is close to resonance with the driving light.

\subsubsection{Basic relations}

For each stochastic realization of
atomic positions, we have an ensemble of $N$ atoms with ground
state $\ket{g}$ and excited state $\ket{e}$, which the incident field
couples via an electric dipole transition with resonance frequency $\omega_0$.
In the low-intensity limit, in which population of the excited state
can be neglected, the expected scattered field can be computed by
treating the atoms as classical linear oscillators whose dipole
moments are given by
\begin{equation}
  \label{eq:dip_def}
  \spvec{d}_j(t) = \unitvec{e} \, \Dc\Pc^{(j)}(t) \textrm{.}
\end{equation}
When analyzing the light and atoms, we adopt the rotating wave approximation where the dynamics is dominated by the frequency of the driving laser
$\omega$. Here, and in the rest of the paper, the light and atomic field amplitudes
refer to the slowly-varying versions of the positive frequency components of the corresponding variables, where the
rapid oscillations $e^{-i\omega t}$ due to the laser frequency have been factored out.
In Eq.~\eqref{eq:dip_def}  $\Pc^{(j)}$ is a  dimensionless dipole amplitude; the index $j=1,\ldots,N$ denotes the $j^{\rm th}$ atom with the position $\stochxv_j$.
This dipole, in turn, emits an electric field with the amplitude
\begin{equation}
  \label{eq:2}
  \epsilon_0\spvec{E}_S^{(j)} (\rv)= \radKernel(\rv - \stochxv_j)\unitvec{e}\,\Dc\Pc^{(j)}\textrm{,}
\end{equation}
where $\radKernel$ is the monochromatic dipole radiation kernel whose
elements are given in Cartesian coordinates by~\cite{Zangwill}
\begin{equation}
  \label{eq:rad_kernel_def}
  \radKernel_{ij}(\spvec{r}) = \left[\frac{\partial}{\partial r_i}
    \frac{\partial}{\partial r_j} - \delta_{ij}\nabla^2\right]
  \frac{e^{ikr}}{4\pi r} -\delta_{ij}\delta(\rv) \textrm{ .}
\end{equation}
There is a characteristic length to the dipole radiation set by the wave number of the light, $k^{-1}=\lambda/2\pi$.

An incident field with
$\cbE_0(\spvec{r},t)$ drives the atomic sample.
The atom $j$ in the ensemble is driven by the electric field
that includes the incident field and the
fields emitted by all other atoms in the system,
\begin{equation}
\spvec{E}_{\rm ext}(\stochxv_j,t) = \cbE_0({\stochxv_j},t) +
\sum_{l\ne j} \spvec{E}_S^{(l)}(\stochxv_j,t).
\end{equation}
This results in the atomic dipole amplitudes satisfying the coupled
dynamics
\begin{eqnarray}
  \label{eq:dynamics_two_level}
  \lefteqn{\frac{d}{dt} \Pc^{(j)}  = \left( i \Delta - \gamma \right)
  \Pc^{(j)} + i\frac{\xi}{{\cal D}} \unitvec{e}^{\ast} \cdot
  \eo \spvec{E}_{\rm ext}(\stochxv_j) \textrm{,} }\nonumber  \\
& &= \left( i \Delta - \gamma \right)
                              \Pc^{(j)} + i\xi \sum_{l\ne j} {\cal G}^{(jl)}
                              \Pc^{(l)} +i \frac{\xi}{{\cal D}}
    \unitvec{e}^{\ast} \cdot \eo\cbE_0(\stochxv_j),\nonumber\\
\end{eqnarray}
with the detuning $\Delta = \omega - \omega_0$, the single-atom
Wigner-Weisskopf linewidth $\gamma$ [Eq.~\eqref{eq:Gamma_E0_def}], and\begin{equation}
\xi={6\pi \gamma\over k^3}\,.
\end{equation}
${\cal G}^{(jl)}$ is the DD coupling between two different atoms $j$ and $l$,
\begin{equation}
  \label{eq:V_2_level_def}
  \mathcal{G}^{(jl)} = \unitvec{e}^\ast \cdot
  \radKernel(\stochxv_j-\stochxv_l) \unitvec{e} \textrm{.}
\end{equation}
The decay rate $\gamma$ may, in fact, be regarded as a result of the dipolar field acting back on the same atom that radiates it.

In typical situations we are mostly interested in the steady-state response. It is then useful to introduce the atomic polarizability
$\alpha=-{\cal D}^2/[\hbar\eo (\Delta+i\gamma)]$
that provides the relationship between the atomic dipole amplitude and the field driving the atom
\beq
\Dc\Pc^{(j)}= \alpha\eo \, \unitvec{e}^\ast \cdot\spvec{E}_{\rm ext} (\stochxv_j) \,.
\eeq

This simulation procedure with coupled dipoles accounts for recurrent
scattering of light between the atoms to all orders, exactly reproducing the dynamics
of light-induced $N$-body correlation functions for stationary atoms in the limit of low light intensity, and was originally introduced for atomic systems
in Ref.~\cite{Javanainen1999a}
(see also~\cite{Lee16}); owing to the simple level
structure and the low light intensity, the coupled-dipole model simulations are  exact even in quantum mechanics~\cite{Javanainen1999a,Lee16}.
The close correspondence between the classical electrodynamics and the quantum-mechanical collective emission of a single-photon excitation for
a cloud of $N$ two-level atoms was also discussed in Ref.~\cite{SVI10}.

The coupled equations for the dipole amplitudes~\eq{eq:dynamics_two_level} are linear, of the form
\begin{equation}
   \dot{\mathrm{b}} = -i \mathcal{H}\mathrm{b} + \mathrm{F} \textrm{,}
 \label{COMPACTEQ}
\end{equation}
where $\mathrm{b}$ is a vector made of the amplitudes $\Pc^{(j)}$, say, $\mathrm{b}=[\Pc^{(1)},\ldots,\Pc^{(N)}]^{\rm T}$, $-i\mathcal{H}$ is the coupling matrix corresponding to the first two terms on the right-hand side of Eq.~\eq{eq:dynamics_two_level}, and $\mathrm{F}$ is a vector whose components
correspond to the driving of the dipoles by the incident
field.

The evolution of the coupled atom-light system can be analyzed using collective eigenmodes $n$ ($n=1,\ldots,N$), which correspond to the
eigenvectors $\mathrm{v}_n$ of $\mathcal{H}$ with the eigenvalues $\delta_n - i
\upsilon_n$. Here $\delta_n$ is the difference between the collective
mode resonance frequency and the resonance frequency of a single, isolated atom, and
$\upsilon_n$ is the collective radiative linewidth.  The matrix $\mathcal{H}$ is in general not Hermitian, hence the nonzero imaginary parts of the eigenvalues.
Moreover, the eigenvectors are not necessarily orthogonal.  In all of our examples, though, they still form a basis. One can therefore uniquely express
the dipole amplitudes and the driving incident field as
\begin{equation}
 \mathrm{b}(t) = \sum_{n} c_n(t) \mathrm{v}_n,\quad  \mathrm{F}(t) = \sum_n
f_n(t) \mathrm{v}_n.
\label{EXPANSION}
\end{equation}
Conversely, given  $\mathrm{b}$ at any time $t$, the first of Eqs.~\eq{EXPANSION} may be construed as a linear set of equations from which to solve the coefficients $c_n$, and likewise for the second equation.

From Eq.~\eqref{COMPACTEQ}, the amplitudes {$c_{n}$} of the collective
mode (here driven at the single atom resonance) satisfy the equation of motion
\begin{equation}
  \label{eq:eqOfMEignModes}
  \dot{c}_n = \left(-i\delta_n - \upsilon_n \right) c_n +
  f_n(t) \, \textrm{.}
\end{equation}
Equivalently, since the atoms are not excited before the incident
field is turned on [$c_n(-\infty)=0$],  the amplitudes satisfy
\begin{equation}
  \label{eq:c_amp_integral}
  c_n(t) = \int_{-\infty}^{t}dt' e^{\left(-i\delta_n -
      \upsilon_n \right)\left(t - t'\right)} f_n(t') \,\textrm{.}
\end{equation}
This is how we handle explicitly time-dependent excitation pulses in the present paper.

For a driving field that is turned on at time $t=0$ and remains constant
in time, the inverse of the collective linewidth $1/\upsilon_n$ is a measure of how
quickly a collective mode will reach its steady state. In the special case of a driving field with a time-independent amplitude and hence  constant $f_n$, the atomic response eventually reaches a steady state. In the present notation the steady state is specified by
\begin{equation}
c_n = \frac{f_n}{i\delta_n+\upsilon_n}.
\label{STEADYRESPONSE}
\end{equation}
The same steady state could be found directly from Eq.~\eq{COMPACTEQ} by solving the linear set of equations $-i \mathcal{H}\mathrm{b} + \mathrm{F} =0$ for the excitation amplitudes $\mathrm{b}$.

Given the vector of the dipole excitations~$\mathrm{b}$, one can compute the electric field at  an arbitrary position $\br$, except for the exact positions of the atoms, from
\begin{equation}
\spvec{E}(\br,t) = \cbE_0(\br,t) +
\sum_{l} \spvec{E}_S^{(l)}(\br,t).
\end{equation}
Of course, the assumption here and in all of our development is that the slowly varying quantities do not change substantially during the time it takes light to propagate across the atomic sample.
That is why the time argument, explicit or implied, is the same throughout our equations. The assumption is valid in light propagation, provided that the group velocity of light does not become
comparable with $L\gamma$ where $L$ denotes the characteristic sample size.

\subsubsection{Inhomogeneous broadening}
\label{sec:voigt}

At the very low temperatures of the experiments the atoms make a homogeneously broadened sample; to the leading order of approximation they do not move at all. We have thus adopted a ``frozen gas" approximation with stationary atoms. In principle one could allow the atoms to move ballistically and in response to the dipole forces between the atoms, even take into account collisions of other origin, and integrate the equations of the polarization amplitudes~\eq{eq:dynamics_two_level} treating the positions $\stochxv_j$ as time dependent. This would make the propagation phases such as $e^{ikr}$ in the dipolar field time dependent, and bring in the usual consequences of the atomic motion such as the Doppler shifts. The drawbacks are that then both our analysis of the collective modes and the notion of a microscopic stationary state break down, and the management of the motion in and of itself greatly complicates the coding.

Instead, we resort to a simple model of inhomogeneous broadening.
Although our aim here is to model the Doppler shifts of moving atoms~\cite{Javanainen2014a}, the simulation procedure for the inhomogeneous broadening is general and applies also
to other resonant emitter systems~\cite{JenkinsRuostekoskiPRB2012b}, such as those consisting of circuit resonators or quantum dots in which case typical
sources for inhomogeneous broadening
are fabrication imperfections.

In order to incorporate the inhomogeneous broadening to the stochastic model we add to the detuning of the $j^{\rm th}$ atom a random quantity $\zeta_j$ that has a Gaussian distribution with the root-mean-square width $\Omega$.
For instance, with $\Omega=ku$ and $u=\sqrt{k_BT/m}$ being the thermal velocity, this would be the standard model for the Doppler broadening of a resonance line of an atom with mass $m$ at the temperature $T$. Instead of the usual Lorentzian resonance line of width $\gamma$ (normalized here to the maximum value of one), we use the Voigt profile (the convolution of a Lorentzian and a Gaussian):
\begin{align}
V(\Delta,\gamma,\Omega)& =  \frac{1}{\sqrt{2\pi}\,\Omega}\int d\zeta \frac{\gamma^2}{(\Delta+\zeta)^2+\gamma^2} e^{-\frac{\zeta^2}{2\Omega^2}} \nonumber\\
=& \sqrt\frac{\pi}{2} \frac{\gamma}{\Omega}\Re\left[
e^{\frac{(\gamma -i \Delta )^2}{2\Omega ^2}}
\text{erfc}\left(\frac{\gamma -i \Delta }{\sqrt{2}\, \Omega }\right)
\right],
\end{align}
where \text{erfc} stands for the complement of the error function. We then fit $A\,V(\Delta-\delta\omega,\gamma,\Omega)$ to a resonance line, with any of $A$, $\delta\omega$, $\gamma$ or $\Omega$ regarded as variable parameters as needed, providing an estimate of the resonance shift $\delta\omega$ of the sample from the atomic resonance $\omega_0$.

\subsubsection{General role of DD interactions}

\begin{figure}
  \centering
  \includegraphics[width=0.6\columnwidth]{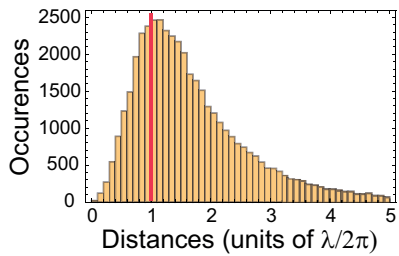}
  \caption{ Stochastic simulation of the probability distribution of the distance between an atom and its nearest neighbor in a cloud of $N = 450$ atoms.
The characteristic interatomic separation distance $1/k$ that represents the length scale below which the light-induced DD interactions become especially strong is highlighted in the plot
by a vertical line.  }
  \label{fig:dipole_dipole_int1}
\end{figure}

To appreciate qualitatively the effect that the DD interactions have on the atomic
response, we display in Fig.~\ref{fig:dipole_dipole_int1} the distribution of nearest-neighbor separations in an atomic gas
of $N=450$ atoms corresponding to the experimental setup. The interatomic separation distance $1/k$ that represents the characteristic length scale associated with the dipole radiation and below which the light-induced DD interactions become especially strong is highlighted in the plot.

\begin{figure}
  \centering
  \includegraphics{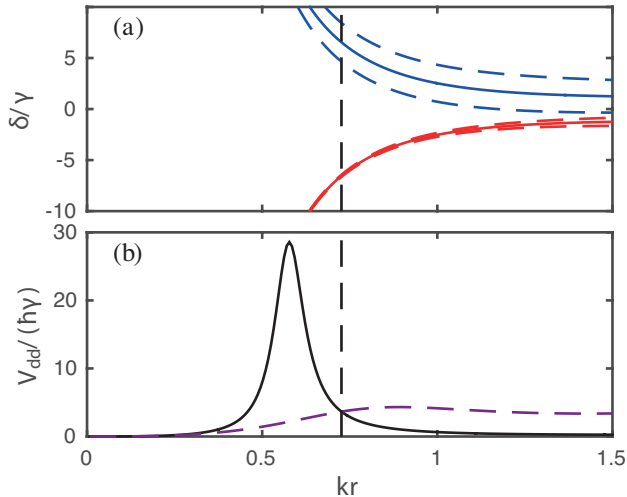}
  \caption{The sensitivity of DD interactions to the
    distance between an atom and its nearest neighbor. Here two two-level atoms are positioned so that their axis of separation is perpendicular to the direction of propagation
    of the driving plane wave and the dipoles are perpendicular to the axis of separation.
    (a) The shift $\delta$ of the superradiant collective mode (blue, solid top curve) and subradiant mode
    (red, solid lower curve) of the two atoms as a function of the distance between them.  The dashed lines that
    bracket the solid lines indicate the collective widths of the
    modes.
    (b) The DD interaction energy $V_{\rm dd}$ as a function
    of the separation distance.
    The two curves are for the driving light tuned to the resonance frequency of the superradiant
    mode of the atoms when separated by the mean distance between the atoms in a cloud of $N = 450$ atoms as in Fig.~\ref{fig:dipole_dipole_int1} (solid line), and to the single-atom
    resonance (dashed line).
  }
  \label{fig:dipole_dipole_int}
\end{figure}

In the case of two atoms, there are two collective eigenmodes for the optical response, each with a distinct frequency and decay rate. The frequencies and decay rates are shown in Fig.~\ref{fig:dipole_dipole_int}(a). As a result of the DD interactions mediated by the electric field, the eigenfrequencies of the nonrelativistic theory diverge in the limit of vanishing distance between the atoms as $1/r^3$. In this same limit the linewidth of one of the modes approaches $2\gamma$, twice the linewidth of an individual atom, and the linewidth of the other mode tends to zero. We call these modes, respectively, superradiant and subradiant~\cite{Dicke1954,GrossHarochePhysRep1982}.

Figure~\ref{fig:dipole_dipole_int}(b) shows the strength of the DD interaction
\begin{eqnarray}
  \label{eq:1}
  V_{\rm dd} &=& -\Dc\Pc^{(1)\ast}\unitvec{e}_y\cdot
  \spvec{E}_S^{(2)}(\stochxv_1) \nonumber \\
  & & - \Dc\Pc^{(2)\ast}\unitvec{e}_y\cdot
  \spvec{E}_S^{(1)}(\stochxv_2)  + {\rm c.c.}
\end{eqnarray}
as a function of the distance between the two atoms for two different fixed tunings of the driving light. The interaction tends to zero at short distances.

Our example presents two important lessons about DD interactions between two atoms. First, as the characteristic frequencies of the collective modes $\delta_n$ diverge at short distance, the external drive will no longer excite the collective modes [c.f.~Eq.~\eq{STEADYRESPONSE}]; nor will the individual atoms be polarized. Moreover, even if the DD interaction formally seems to diverge as $1/r^3$ at short distances, the decoupling of the atoms from the driving light wins out. One might surmise that at very short distances the DD interaction between two atoms becomes a calamity. Not so: Both atoms drop out of consideration altogether. Second, when the characteristic distance between the atoms is comparable to $1/k$, their interactions with light depends sensitively on their relative positions. Local density fluctuations are important, and the conventional electrodynamics for a many-atom system that assumes that the density of the atoms is sufficient to characterize their interactions with light is liable to break down.

\subsection{Coupled dynamics of $^{87}$Rb atoms}
\label{sec:87rb-scattering}

The multilevel structure of alkali-metal atoms, such as $^{87}$Rb
complicates, the collective dynamics of a cold gas with respect to
that of two-level gases~\cite{Pellegrino2014a,Jenkins_thermshift}.
Rather than atoms beginning in a single ground state, they could in
principle occupy any linear combination of Zeeman ground states before
the incident field interacts with the ensemble.
Furthermore, different polarizations of light interact with the
various transitions between Zeeman states in the ground and hyperfine
levels in different ways according to the corresponding Clebsch-Gordan
coefficients.
Here, we simulate the response of an atomic
cloud whose atoms are in an incoherent mixture of Zeeman levels.
We will then discuss the dynamics of an individual stochastic
realization, and examine the distributions of collective resonance
frequencies and decay rates that appear in a dense cloud of atoms.

\subsubsection{Dealing with Zeeman states}

In the experiments the optical pumping process prepares each atom in
an incoherent mixture of Zeeman states $\ket{g,M}$ with probabilities
$p_M$ ($\sum_M p_M = 1$), where
$M=-g,\ldots, g$ indicates the magnetic quantum number of a state in
the hyperfine level $g$.
To account for this incoherent mixture in our stochastic simulations,
for each realization of atomic positions
$\{\stochxv_1,\ldots,\stochxv_N\}$ we also sample a magnetic quantum
number $M_j$ from each atom $j$ according to the probability
distribution $p_M$. In the discussion below
the ground-state magnetic quantum numbers $M_j$ are constants
characteristic of one particular realization of the $N$-atom sample.

The EM response of each atom is then
characterized by three amplitudes $\Pc_{M_j,M_j+\sigma}^{(j)}$, where
$\sigma = -1,0,1$ is an index indicating the polarization, and
$\Pc_{M_j,\eta}^{(j)}$ is an amplitude corresponding to the
$\ket{g,M_j}_j \leftrightarrow \ket{e,\eta}_j$ transition.  In the low
light-intensity limit, the electric dipole moment is
\begin{equation}
  \label{eq:dip_with_degeneracy}
  \spvec{d}_j(t) = \Dc \sum_{\sigma=-1}^{1}
  \cgshort{M_j}{\eta}{\sigma} \pola_{\sigma}\Pc_{M_j,\eta}^{(j)}(t)  \,\textrm{,}
\end{equation}
where $\cgshort{M}{\eta}{\sigma} \equiv \langle
F_e\eta;1F_g|F_gM;1\sigma\rangle$ are Clebsch-Gordan coefficients
for the corresponding dipole transition ($F_f$ is the total atomic
angular momentum of hyperfine level $f$), and the polarization vectors are defined in Eq.~\eq{polarizationvec}.
We use Greek indices to denote the Zeeman states of the excited level $e$, and adopt the convention that repeated Greek indices are summed over.
Here we consider the case $F_g=2$ and $F_e=3$.

As with two-level atoms, each $^{87}$Rb atom $j$ in our system is
driven by the field $\spvec{E}_{\rm ext} (\stochxv_j,t)$, which comprises the incident
field $\cbE_0$ and the scattered fields $\spvec{E}_S^{(l)} $ ($l
\ne j$). As a result, the amplitudes of the atomic dipoles associated to a given transition have the dynamics
\begin{eqnarray}
  \label{eq:3}
  \frac{d}{dt} \Pc_{M_j\eta}^{(j)}& = &\left(i\Delta_{M_j\eta}
  - \gamma\right) \Pc_{M_j\eta}^{(j)} \nonumber \\
                                         & &
                                             +i\xi \sum_{l\ne j}
                                             \cgshort{M_j}{\eta}{\sigma}
                                             \mathcal{G}_{\sigma\varsigma}^{(jl)}
                                             \cgshort{M_l}{\zeta}{\varsigma}
                                             \Pc_{M_l\zeta}^{(l)} \nonumber\\
                                         & &   +i \frac{\xi}{{\cal D}}
                                             \cgshort{M_j}{\eta}{\sigma}
                                             \pola_{\sigma}^* \cdot \eo
                                             \cbE_0(\rv,t) \text{,}
\end{eqnarray}
where the amplitudes are coupled through the DD
interaction between dipoles of orientation $\pola_\sigma$ and
$\pola_\varsigma$,
\begin{equation}
  \label{eq:V_multi_level_def}
  \mathcal{G}^{(jl)}_{\sigma\varsigma} = \unitvec{e}^\ast_\sigma \cdot
  \radKernel(\stochxv_j-\stochxv_l) \unitvec{e}_{\varsigma} .
\end{equation}
In addition to the incident and scattered electric fields, the applied
$B=1\,\mathrm{G}$ bias field in the $x$ direction shifts the individual Zeeman levels.
The
resonance frequency of the $\ket{g,M} \leftrightarrow \ket{e,\eta}$
transition is shifted so that the incident field is detuned from that
transition by
\begin{equation}
\Delta_{M\eta} \equiv \omega - \big[\omega_0 +
\frac{\mu_BB}{\hbar} (g_eM - g_g\nu)\big]\,,
\end{equation}
where the Land\'{e} g-factors are
$g_g\approx 0.50$ and
$g_e\approx 0.67$, and
$\omega_0$ is the resonance frequency unperturbed by the magnetic bias
field.
The total electric field elsewhere except at the position of the atoms is
\begin{equation}
 \spvec{E}(\br,t) =  \cbE_0(\br,t) + \frac{\Dc}{\epsilon_0}\sum_{j }
  \cgshort{M_j}{\eta}{\sigma} \radKernel(\rv - \stochxv_j)\pola_{\sigma}\Pc_{M_j,\eta}^{(j)}(t)  \,.
\label{RBFIELD}
\end{equation}
The steady-state response then follows from
\begin{equation}
 {\cal D} \Pc_{M_j\eta}^{(j)} =  \alpha_{M_j\eta} \sum_\sigma \cgshort{M_j}{\eta}{\sigma}
     \pola_{\sigma}^* \cdot \eo\spvec{E}_{\rm ext} (\stochxv_j,t)
  \,,
\end{equation}
where the polarizability now reads
\beq
\alpha_{M_j\eta}=-{{\cal D}^2\over \hbar\eo (\Delta_{M_j\eta}+i\gamma)}\,.
\eeq

The construction of the collective modes and their use in the analysis of the response of the system proceeds in the same way as with the two-level system, except that the vector of polarization amplitudes $ \mathrm{b}$ for $N$ atoms is now made of the $3N$ polarization amplitudes $\Pc_{M_j\eta}^{(j)}$ with $\eta=M_j-1,M_j,M_j+1$. In our analysis of the
Rb cloud we will also introduce an inhomogeneous broadening for the resonance frequencies similarly as it was explained in the case of two-level atoms.

The numerical results shown in this paper have been tested by two independently developed sets of codes. The codes are based on a collection of C++ classes capable of handling an arbitrary angular momentum degenerate multilevel atom dipole-coupled to an arbitrary set of driving fields, etc. The same numerical methods
have also been applied to solving collective electromagnetic response of plasmonic and microwave resonator arrays~\cite{JenkinsLongPRB,AdamoEtAlPRL2012,CAIT}, with appropriate extensions to include the magnetic properties
of these materials.

\section{Numerical and experimental results}
\label{sec:results}

In this section we present numerical simulation results and experimental findings of the optical response of an atomic ensemble in an elongated, ellipsoidal trap. The experimental setup is explained in Sec.~\ref{Sec:Exp_setup} and concerns a cold $^{87}$Rb vapor with several internal electronic ground states participating in the light scattering processes. In numerical simulations we also compare the cold-atom simulations with those of a thermal vapor. We begin by analyzing some generic collective properties of the atomic ensemble by calculating the collective radiative excitations modes of the atoms and the corresponding collective radiative resonance linewidths and line shifts.

\subsection{Cooperative modes}
\label{sec:cooperative-modes}

The collective mode characteristics for a particular realization of
atomic positions strongly influence the response of the
ensemble as a whole.
As seen in Eqs.~\eqref{eq:eqOfMEignModes} and
\eqref{eq:c_amp_integral}, the closer the collective resonance
frequencies and decay rates are to those of a single atom, the better
the scattering dynamics can be approximated by independent atoms.
A broad distribution of resonance frequencies or decay rates, as
will be evidenced in Secs.~\ref{Sec:Exp_results} and
\ref{sec:optic-resp-incid}, alters the optical response.

\subsubsection{Eigenmodes}

\begin{figure}
  \centering
  \includegraphics[width=0.98\columnwidth]{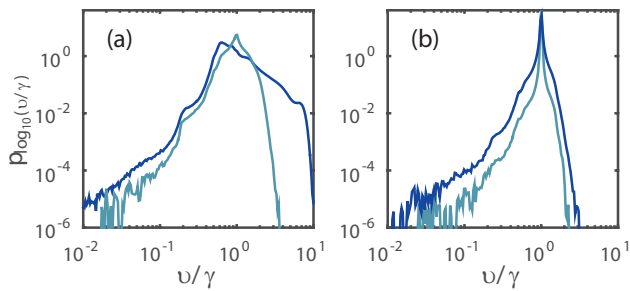}
  \caption{Distribution of collective mode decay rates in a cloud of
    homogeneously broadened (cold) $^{87}$Rb atoms (a), and in an
    inhomogeneously broadened (hot) cloud in which the single-atom resonance
    frequencies have a Gaussian distribution with the root-mean-square
    width $\Omega=100\gamma$ (b).  The samples contain $N=50$
    (light bottom curve) and $N=450$ (dark top curve) atoms.  The distributions were computed
    from a histogram of the values of $\log_{10}(\upsilon_n)$ generated
    by $51200$ samples of atomic positions in which the Zeeman states
    have initial populations $p_0=p_1=p_2 = 1/3$ and
    $p_{-1}=p_{-2}=0$. }
  \label{fig:hist_gamma_hom}
\end{figure}

\begin{figure}
  \centering
  \includegraphics{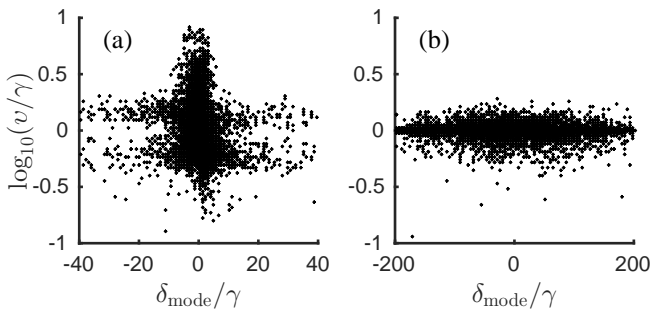}
  \caption{Representation in the complex plane of the eigenvalues associated with the collective modes for a particular realization of the atomic positions. Each mode is represented by a point, with  the coordinates being the
mode resonance frequencies $\delta_{n}$ and
   its decay rates $\gamma_{n}$.  (a)
    Homogeneously  and (b) inhomogeneously broadened (with
    a root-mean-square Gaussian width of $100\gamma$) samples of
    450 $^{87}$Rb atoms. }
\label{fig:hist_delta_hom}
\end{figure}

Figure \ref{fig:hist_gamma_hom} shows the distribution of the
logarithm of collective mode decay rates $\log_{10} (\upsilon /
\gamma)$ for cold (homogeneously broadened) and thermal (inhomogeneously broadened) gases  of $^{87}$Rb atoms with $N=450$.
This distribution represents the probability density of obtaining a
particular value of $\log_{10}(\upsilon/\gamma)$ on randomly selecting
a collective mode from a gas with randomly chosen atomic positions.
We computed the distribution numerically from $51200$ realizations of
atomic positions.

In a cold gas the number of atoms in the ensemble strongly influences the width of the distribution of the collective decay rates.
The system may support collective modes with subradiant and superradiant decay rates spanning
several orders of magnitude.
For $N=50$ and $N=450$ cold $^{87}$Rb atoms, one percent of collective modes have decay rates of less than $0.45\gamma$ and $0.39\gamma$, respectively, and the median linewidths in these samples are $0.98\gamma$ and $0.79\gamma$, respectively.
The inhibition of the light-mediated interactions in thermal ensembles is also shown in the distribution of the decay rates that is notably narrower. For both $N=50$ and $N=450$ atoms, the median linewidth matches that of a single atom, while one percent of the collective decay rates are below $0.90\gamma$ and $0.62\gamma$, respectively. Overall, we find that increasing the density of cold atoms makes the median value of the linewidth smaller, and generates a long tail of subradiant mode decay rates.

One can also see how collective modes differ in cold and
thermal gases by examining the joint distributions of $\delta_{n}$ and
$\log_{10}\upsilon_{n}$.
Scatter plots of $\delta_n$ and $\log_{10}\upsilon_{n}$ obtained from five
realization of atomic positions in samples of 450 atoms are shown in
Fig.~\ref{fig:hist_delta_hom}.
In a cold gas, many of the mode resonance frequencies are shifted
from the single-atom resonance, while collective decay rates span
several orders of magnitude.
The eigenmodes that exhibit larger resonance shifts also tend to be either fairly subradiant or superradiant and do not have linewidths that
are close to the linewidth of a single isolated atom, indicating that large resonance shifts are correlated with the changes in the decay rates.
Interestingly, the largest shifts are not associated with the most subradiant or superradiant modes.

The distribution of mode resonance frequencies in the inhomogeneously
broadened sample, on the other hand, mostly reflects the Gaussian
distributions of single-atom resonance frequencies. Furthermore, the
collective mode decay rates closely match single-atom decay rates,
indicating that the atoms respond to light nearly independently.

\begin{figure}
  \centering
  \includegraphics[width=0.95\columnwidth]{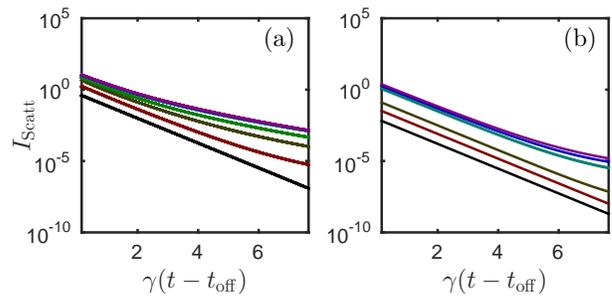}
  \caption{ The simulated scattered power collected by the lens as a function of time
    after a $125$ns square excitation pulse is turned off for (a) a
    homogeneously broadened sample, and (b) an inhomogeneously broadened
    sample of $^{87}$Rb atoms with the Doppler broadening of $100\gamma$.
    Here the atoms are trapped in a thermal equilibrium during and after the pulse.
    We show the intensity scattered into the detection apparatus (the curves from bottom to top) for the atom numbers $N=1$ (black), $N=5$ (red), $N=20$ (brown),
    $N=50$ (green), $N=200$ (cyan), $N=325$ (blue), and $N=450$ (purple).  The light is on single-atom resonance.
    The varying
    rates of decay in the homogeneously broadened sample indicate the
    participation of both superradiant and subradiant collective
    modes, while the constant decay rates of the inhomogeneously
    broadened samples at the lowest atom numbers indicate that the atoms are radiating independently.
  }
  \label{fig:intens_after_pulse}
\end{figure}

\subsubsection{The decay of collective modes}

The role different collective decay rates play in the optical response
of the ensemble can be inferred by calculating the temporal profile of
scattered light.
Consider a $125$ns square pulse, tuned to the frequency that, on average, most
strongly scatters from a single $^{87}$Rb atom.  Here, for simplicity,
we neglect the rise and fall times.
After the pulse is turned off, the collective modes with higher
decay rates contribute most strongly to light emission.  But later,
when more subradiant modes are present in the ensemble, the reduction
in scattered light intensity slows with time, reflecting the
excitation of subradiant modes that live longer.
This behavior is illustrated for samples of various numbers of atoms
in Fig.~\ref{fig:intens_after_pulse}.
Samples with more atoms, and hence broader distribution of decay
rates, produce scattered radiation with a temporal profile that decays
more slowly as time goes on.  The deviations of the curves from straight lines
in Fig.~\ref{fig:intens_after_pulse} indicate the co-existence of different exponential
decay rates in the temporal profile. These represent non-negligible atom populations
of different collective modes that exhibit different linewidths.
In a thermal gas, where inhomogeneous
broadening weakens DD interactions, this effect of subradiant modes requires
higher densities to manifest itself than in a cold gas.

Using a cold and low-density, but optically thick, atom vapor the effects of subradiance were recently observed~\cite{Guerin_subr16}.
In these experiments the long tails of the decay distribution indicated the existence  of
subradiant mode excitations.
Our simulations show that the signal in a dense gas could be notably enhanced.

\subsection{Experimental measurements of scattered light}\label{Sec:Exp_results}

The  experimental setup was described in Sec.~\ref{Sec:Exp_setup}.
Reference~\cite{Pellegrino2014a} reported measurements of light
scattering performed by sending a series of light pulses on the atomic cloud,
hereafter referred to as the ``burst excitation'' method. Additional experimental protocols based on imaging with a single laser pulse were reported in Ref.~\cite{Jenkins_thermshift}.
These were used to rule out potential systematics and check the robustness of the resonance shift measurements of Ref.~\cite{Pellegrino2014a}.

\subsubsection{ Burst excitation method}

In the first experiments described in detail in Ref.~\cite{Pellegrino2014a}, we excited the cloud
with 125~ns long pulses with flat-top temporal profile
(rise time 7 ns) while the cloud had been released
in free flight for 500 ns. The switching-off of the dipole trap was meant
to eliminate light shifts caused by the
trapping laser. The atomic density was nearly frozen during this period. However,
as the amount of scattered light collected in a single realization of the experiment was
very small, corresponding to less than $10^{-3}$ photons, we recycled the cloud many times
and  repeated sequences of
excitation and recapture  two hundred times with the same cloud of
atoms.
Excitation pulses were thus interleaved with $1\,\mathrm{\mu s}$ periods of
recapture in the dipole trap.
Further runs were then performed with newly prepared atomic clouds.
The duration of the excitation pulse of $T=2.4/\gamma=125$ ns
results from a compromise: it
allows the
atoms to approach the steady state, while minimizing the heating of the atoms.
The choice of the number of sequences of excitation and recapture is then a
trade-off between getting a good signal-to-noise ratio and avoiding
light assisted losses~\cite{Fuhrmanek2012} or heating of the cloud,
as both effects would lower the cloud's density.
We have checked that the temperature of the cloud did not rise by more than $5\%$ after the
entire set of pulses and that less than $5\%$ of the atoms were depumped in
the $(5S_{1/2},F=1)$ hyperfine level.

For each excitation pulse, we recorded the total time-integrated power
of the light scattered by the cloud arriving at the CCD camera, as
shown in Fig.~\ref{fig:setup}.
For each number of atoms, we monitored the amount of scattered light as a function of the detuning
of the excitation laser. As shown in Ref.~\cite{Pellegrino2014a}, the lines are well fit by a Lorentzian. From the fit we extracted the amplitude $A$, the shift of the center of the line $\delta\omega_{\rm c}$,
as well as the HWHM $\gamma_{\rm c}$.
The results for the shift and the width are plotted in
Figs.~\ref{fig:Exp_shift} and~\ref{fig:Exp_width} as a function of the peak atomic density $\rho$ of the cloud. The latter is calculated from
the measurement of the number of atoms,  the temperature, and the knowledge of the trapping potential.
Each point, for a given atom number, corresponds to an average over typically $1000$ newly-loaded clouds.
We observe a small red-shift ($|\delta\omega_{\rm c}|\lesssim 0.3\gamma)$
and a broadening of the line, showing a sharp increase with the density for $\rho\lesssim 10^{14}$ at/$\rm cm^3$.

\begin{figure}
\includegraphics[width=\columnwidth]{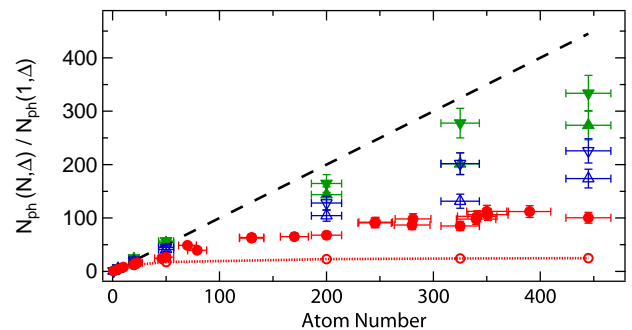}
\caption{Amount of scattered photons $N_{\rm ph}(N,\Delta)$ detected in the $z$ direction, versus the number of atoms $N$, for different detunings $\Delta$ of the laser. The number of detected photons for each atom number is normalized to  the single atom case, $N_{\rm ph}(1,\Delta)$, at the same detuning. The detunings are $\Delta=0$ (red circle), $\Delta=\pm2\gamma$ (up/down open triangles), and $\Delta=\pm5\gamma$  (up/down filled triangles). Red line: result of the numerical simulation of a homogeneously-broadened gas. }
\label{Fig:suppression_fluo}
\end{figure}

Let us consider now an incident field tuned to the frequency that maximizes the
scattered intensity from a single atom.
If $N$ atoms were to scatter independently, the fluorescence of the
ensemble would be $N$ times that of a single atom.
We find that the light scattered in the $z$ direction at resonance does not
increase linearly with the number of atoms as one would
expect for noninteracting atoms, but actually increases
more slowly. It is shown in Fig.~\ref{Fig:suppression_fluo} that this is also the case
off resonance. We observe that the amount of scattered light is strongly suppressed on resonance as
the number of atoms increases, and that we gradually recover the behavior of noninteracting atoms as we detune
the laser away from resonance.

\subsubsection{Complementary protocols}

\begin{figure}
\centering
\includegraphics[width=\columnwidth]{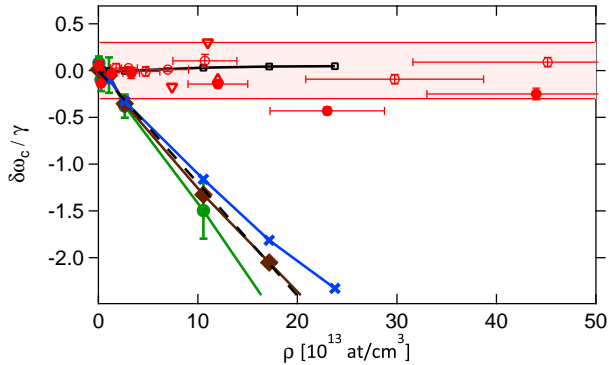}
\vspace{-0.7cm}
\caption{Line shift as a function of the atom density. Filled red circles: excitation with bursts of $200$ light pulses;
each point corresponds to a different number of atoms. Upper and lower red triangles: excitation with respectively
75 and 1 pulses per burst. Empty red circles: excitation with one pulse after a variable time of flight; the cloud contains $\sim 450$ atoms. The shaded area indicates the laser linewidth of $\pm0.3\gamma$. The error bars on the densities result from the cumulated measurement uncertainties on the trap size ($\sim 10\%$), atom number ($\sim 10\%$) and temperature ($\sim 10\%$). Decreasing values of the density correspond to time-of-flights $\Delta t =(0.7, 1.7, 2.7, 3.7, 4.7, 6.7, 8.7, 20.7)\mu$s.
Error bars on the experimental shift are from the fit of the fluorescence resonance spectrum by a Lorentz function.
Numerical simulations: Shift of the line for homogeneously (black empty squares) and inhomogeneously broadened samples with root-mean-square spectral broadening of $10\gamma$ (blue crosses), $20\gamma$ (brown diamonds), and $100\gamma$ (green circles). Error bars: $95\%$ confidence intervals on the shift obtained from the fit of the spectrum to the Voigt profile (see text). Dashed line: estimated Lorentz-Lorenz shift.}
\label{fig:Exp_shift}
\end{figure}

\begin{figure}
\centering
\includegraphics[width=\columnwidth]{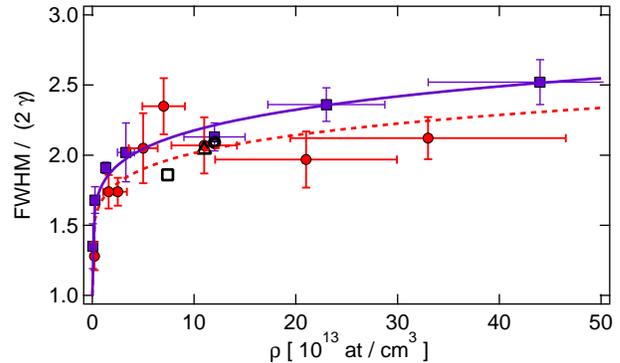}
\caption{ Full width at half maximum of the lines  from the experiments where many bursts are sent onto the cloud (purple squares), and from probing the cloud after a time-of-flight measurement (red circles). Also shown are the results of the burst experiment for different number of pulses sent onto the cloud. The open square correspond to 350 atoms, at 190 $\mu$K, probed with a single pulse. Open circle: 350 atoms at 140 $\mu$K, one pulse. Open triangle: 340 atoms at 150 $\mu$K, 75 pulses. Error bars on the densities: same as in Fig.~\ref{fig:Exp_shift}. Error bars on the width are from the fit of the fluorescence spectrum to a Lorentz function. The solid and dashed lines are phenomenological fits by a power law to the burst and time-of-flight method data, respectively. }
\label{fig:Exp_width}
\end{figure}

To rule out possible systematics due to the repetition of excitation pulses, we performed complementary measurements using the protocol described above~\cite{Jenkins_thermshift} but where we reduced the number of pulses per burst~\footnote{The pulse length is then increased to $700$\,ns to keep the integration time reasonable.}. The results, which we report in Figs.~\ref{fig:Exp_shift} and~\ref{fig:Exp_width}, do not indicate any significant change.
While this does not entirely exclude the possibility of variations of the density during a single pulse, it does rule out the possible cumulative effect from sending several pulses on the same cloud.
Finally, we performed measurements with excitation intensities at even lower levels (down to $I/I_{\rm sat} =0.001$). We still did not see any significant shift in the resonance.

In order to check for the robustness of the absence of the shift in a cold atomic sample, we also implemented a new protocol for the excitation.
Instead of varying the density by modifying the atom number we change the geometry of the cloud. After having trapped $\sim 450$ atoms, we switched off the trap and varied the free-flight period $\Delta t$ during which the density of the atoms drops as $N/[(2\pi)^{3/2}\sigma_x \sigma_y \sigma_z]$, with $\sigma_i^2(\Delta t)=\sigma_i^2(0)+k_B T (\Delta t)^2/m$ ($i=x,y,z$), and the aspect ratio of the cloud evolves from a highly elongated cigar-shaped cloud to a spherical cloud. We then imaged the atoms with a $2\,\mu$s pulse at a given detuning and repeated the experiment $\sim 1000$ times using a new cloud each time.
The results for the resonance shifts  and widths are shown in Figs.~\ref{fig:Exp_shift} and~\ref{fig:Exp_width} as a function of the peak density of the cloud at the beginning of the excitation
pulse~\footnote{Note that the density drops during the $2\,\mu$s excitation pulse and that this drop is particularly significant for short free-flights.}. The density is  again deduced from
the independent measurements of the trap size, atom number, and temperature of the cloud.
Importantly, the various experimental protocols were implemented over a period of several months and with numerous adjustments to the experimental apparatus,
but the results consistently indicate a very small resonance shift, and a broadening
of the line with the same general shape.

\begin{figure}
\centering
 \includegraphics[width=\columnwidth]{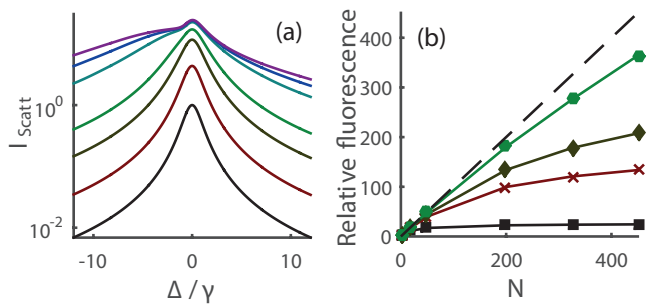}
\caption{(a) Scattered intensity from an ensemble of
 $^{87}$Rb  atoms as a function of detuning
  $\Delta$.  Light is scattered from an ensemble (the curves from bottom to top) with $N=1$ (black), $N=5$ (red), $N=20$ (brown), $N=50$ (green), $N=200$ (cyan), $N=325$, and $N=450$ (purple) atoms.
 (b) relative fluorescence  as a
    function of number of atoms for homogeneously (black squares) and
    inhomogeneously  broadened samples with
    root-mean-square spectral broadening of $10\gamma$ (red crosses),
    $20\gamma$ (brown diamonds), and $100\gamma$ (green circles). The dashed line denotes the result for noninteracting atoms.
    }
\label{fig:partial_polarized_response}
\end{figure}

\begin{figure}
  \centering
  \includegraphics[width=0.75\columnwidth]{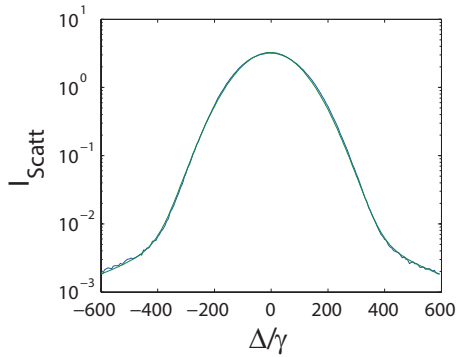}
  \caption{The light scattered from an inhomogeneously broadened
    ($100\gamma$) sample of 450 atoms in the presence of a $1G$
    magnetic field as a function of driving frequency.
    The scattered intensity calculated from
    numerical simulations and the fit to the
    Voigt profile are virtually indistinguishable.
  }
  \label{fig:I_scatt_inhom_fit}
\end{figure}

\subsection{Simulations of the optical response}
\label{sec:optic-resp-incid}

The basic procedures of the numerical simulations of the multilevel ${}^{87}$Rb experiment are described in Sec.~\ref{sec:87rb-scattering} where we model the
experimental setup described in Sec.~\ref{sec:experimental-setup}.
The inhomogeneous broadening is introduced using the techniques explained in Sec.~\ref{sec:voigt}.
From the response of the atomic dipoles to the incident field, we then
calculate the intensity scattered into the detection apparatus as
in the experimental configuration: the lens with numerical aperture
$0.5$ gathers light in the far field centered on the $-z$ direction, and the signal passes through a polarizer rotated
about $-\unitvec{e}_z$ by $55^\circ$ from the $x$ axis.

\subsubsection{Fluorescence spectrum}

Figure~\ref{fig:partial_polarized_response} shows the integrated
scattered intensity as a function of
$\Delta  = \omega -\omega_0'$ for numbers of atoms $N=1$-$450$,
where $\omega_0'$ is the frequency at which a single $^{87}$Rb atom
scatters the most incident light in the presence of the magnetic field.
Calculations on few-atom cold ensembles produce the expected Lorentzian line shapes for the spectra of the scattered intensity.
As $N$ increases, the spectral response begins to deviate from the independent atom scattering both in width and
in the power of scattered light~\cite{Pellegrino2014a}.  The spectral response in the numerics becomes both broader
and more asymmetric, with a fat tail that drops off more slowly for red-detuned incident fields than for
blue detuning.
As the numerically calculated spectrum cannot be modeled by
a Lorentzian, extracting the value of the width is difficult.
More quantitative comparisons between the theory and the measurements
may be obtained from (1) the shift of the resonance and (2) the suppression of the computed peak
fluorescence per atom relative to that of a single atom.

\subsubsection{Resonance shifts}

Figure~\ref{fig:Exp_shift} shows the experimental shifts, which are deduced from Lorentzian fits to the measured spectra, together with the shifts deduced from the
coupled-dipole simulations by simply taking the detunings corresponding to the maxima of the fluorescence intensity. Both shifts are negligible over the range of densities explored here, and are generally smaller than the linewidth of the laser $\sim \gamma/3$ in the experiments.

By contrast, in the simulations we may also introduce the Doppler broadening associated with the thermal Maxwell-Boltzmann distribution of the atomic velocities, and find the resonance shift by fitting to a Voigt lineshape. We find notably larger shifts~\cite{Jenkins_thermshift}. For the Doppler width of $10\gamma$, corresponding to the temperature of $5.5$\,K, the shift is, e.g., at the density $\rho=2.4\times10^{14}$\,cm$^{-3}$, already $50$ times larger than the stationary-atom result. Increasing the Doppler broadening further has a weaker effect on the shift. The quality of the fit of the Doppler-broadened lineshape to the Voigt profile is illustrated in Fig.~\ref{fig:I_scatt_inhom_fit}.

In continuous effective-medium electrodynamics a natural energy scale for the resonance shifts is the Lorentz-Lorenz (LL) shift $\Delta_{\rm LL} = -2\pi\gamma\rho/k^3$,
and at low atom densities $\rho$ one
expects a shift of a resonance $\propto \rho/k^3$ also from dimensional analysis. We may estimate the LL shift by $\rho$ at the center of the trap
(dashed line in Fig.~\ref{fig:Exp_shift}). We find that the LL shift is absent both in the experiments and in the electrodynamics simulations of
a cold gas. By contrast, introducing inhomogeneous broadening restores a resonance shift that is roughly equal to the LL shift $\Delta_{\rm LL}$,
as illustrated in Fig.~\ref{fig:Exp_shift}.

The difference between the optical responses of the cold and thermal atomic ensembles may be understood by the change in the light-induced
DD interactions.
With increasing inhomogeneous broadening the atoms are simply farther away from resonance with the light sent by the other atoms, which reduces the
light-mediated interactions~\cite{Javanainen2014a}. Moreover,
the response of a cold, dense vapor is characterized by the many-atom collective excitation modes. In our case (this generally depends on the geometry of
the sample and the excitation protocol~\footnote{The shift can be recovered in the low atom densities or when a specific collective mode that exhibits a shift
is driven~\cite{JavanainenMFT,Roof16,Araujo16}}) the highly excited modes exhibit resonance frequencies close to the single atom resonance, and the shift in
the observed spectrum consequently is small.
In contrast, in thermal ensembles the shift can be qualitatively attributed to the standard local-field corrections~\cite{Jackson,BOR99} that give rise to the LL shift.

\subsubsection{Scattered intensity}

In addition to the absence of the resonance shift, the collective response
suppresses the resonant fluorescence.
The numerical calculations predict the suppression of the scattered light when the number of
atoms increases~\cite{Fuhrmanek2012,Pellegrino2014a}. The homogeneously broadened case, as shown in Fig.~\ref{Fig:suppression_fluo}, is in good agreement
with the experimental data for $N\alt 50$. For $N\agt 50$, the
agreement is only qualitative, as the effects of the dipole-dipole interactions are found to
be less pronounced experimentally.

By contrast, as shown in Fig.~\ref{fig:partial_polarized_response}(b),
adding a small amount of inhomogeneous broadening reduces the suppression of the scattering.
The fluorescence of the inhomogeneously-broadened samples eventually also approaches
the ideal limit of noninteracting atoms as the broadening reduces the interactions.

\subsubsection{Finite pulse duration}

The results shown thus far indicate how the ensemble of atoms would
scatter in the steady state when exposed to a monochromatic driving
field of frequency $\omega$.
In the experiments, on the other hand, the atoms are excited by a
pulse with a carrier frequency $\omega$ that is about $125$~ns long. The apparatus then captures the scattered intensity integrated over time for each pulse.
The spectral width of the driving pulse would tend to smoothen out some
of the features that appear in the steady-state responses, since the time integrated response is proportional to a convolution of the pulse spectrum with the steady-state spectral response.

\begin{figure}
  \centering
  \includegraphics[width=0.45\columnwidth]{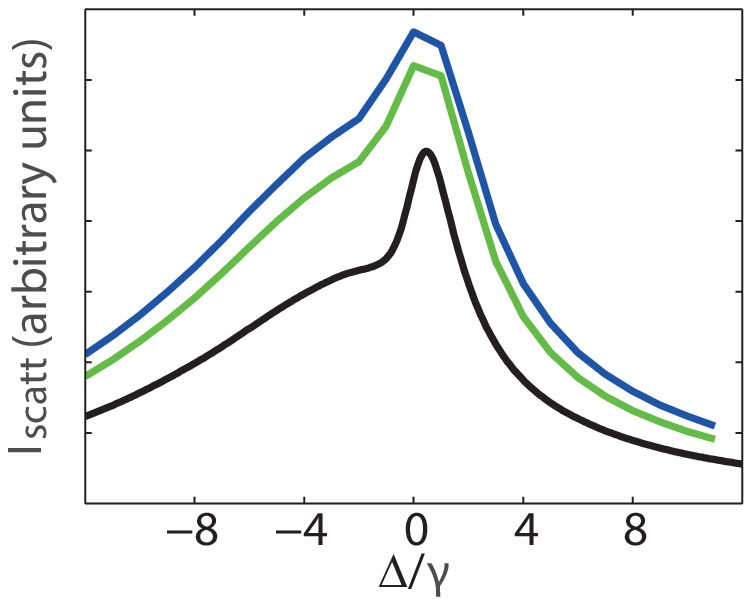}
  \includegraphics[width=0.51\columnwidth]{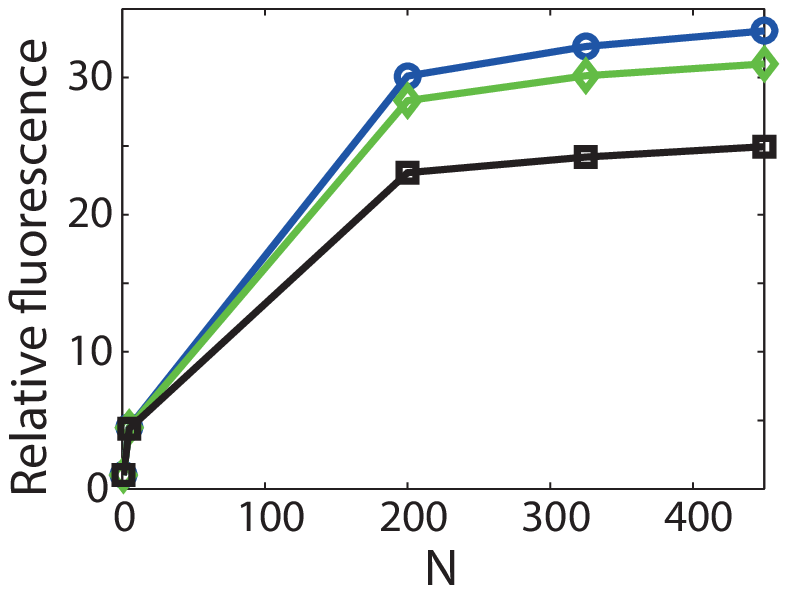}
  \caption{ Differences between the responses to the steady-state and pulsed laser excitations.
  Left: The time integrated intensity scattered from $N=450$ atoms.
  Each curve is normalized
    to the peak time-integrated intensities scattered from a single
    atom driven by the incident field corresponding to that curve. Right:
  The peak relative fluorescence (at the frequency that
    maximizes scattering from a single atom) of the time integrated intensity as it
    depends on the number of atoms.
Curves from bottom to top: the steady-state response (black),  a smoothed square pulse
    [Eq.~\eqref{eq:smoothe_pulse}] (green), and a square [Eq.~\eqref{eq:Square_pulse}] (blue).
  }
  \label{fig:Integrated_Intens_square_smothe_steady_compare}
\end{figure}

Figure~\ref{fig:Integrated_Intens_square_smothe_steady_compare} compares the response of
ensembles of $^{87}$Rb atoms to two pulse profiles and to the steady-state illumination. One pulse
is a square pulse with the temporal
profile
\begin{equation}
  \label{eq:Square_pulse}
  \spvec{E}_{\mathrm{sq}}(t) = E_0 \unitvec{e}_{\mathrm{in}}
  \Theta(t)\Theta(T-t) \, \textrm{,}
\end{equation}
where $T$ is $125$ ns.
The second is a smoothed square pulse that closely resembles the experimental pulse profile,
  \begin{equation}
    \label{eq:smoothe_pulse}
    \spvec{E}_{\mathrm{sm}}(t) = E_0 \sqrt{\frac{1}{1+\exp\left(-\frac{t-t_{\mathrm{on}}}{\Delta{}t}\right)}
      -
      \frac{1}{1+\exp\left(-\frac{t-T-t_{\mathrm{on}}}{\Delta{}t}\right)}} \,\textrm{;}
  \end{equation}
here $T \simeq 124$ns and $\Delta t \simeq 7.56$ns is the pulse
rise/fall time.

The scattered intensities shown in
Fig.~\ref{fig:Integrated_Intens_square_smothe_steady_compare} are
normalized to the peak single-atom integrated intensity for the
respective pulse shape.
Since the smoothed pulse is spectrally narrower than the square pulse,
the resonance width is consistently slightly narrower than
that of the square pulse. Both pulsed excitations, however, are broader than the steady-state response.
The relative fluorescence of the smoothed pulse is also slightly
smaller than that of the square pulse, but not suppressed as much as
that of the steady state.

\subsubsection{Ground-state populations and Zeeman splitting}

In the simulations discussed until now the populations of the initial state of the Rb atoms were always $p_0=p_1=p_2=1/3$. Our next topic about the simulations is the question about
the sensitivity of the results to the initial state of the atom.

\begin{figure}
  \centering
  \includegraphics[width=0.48\columnwidth]{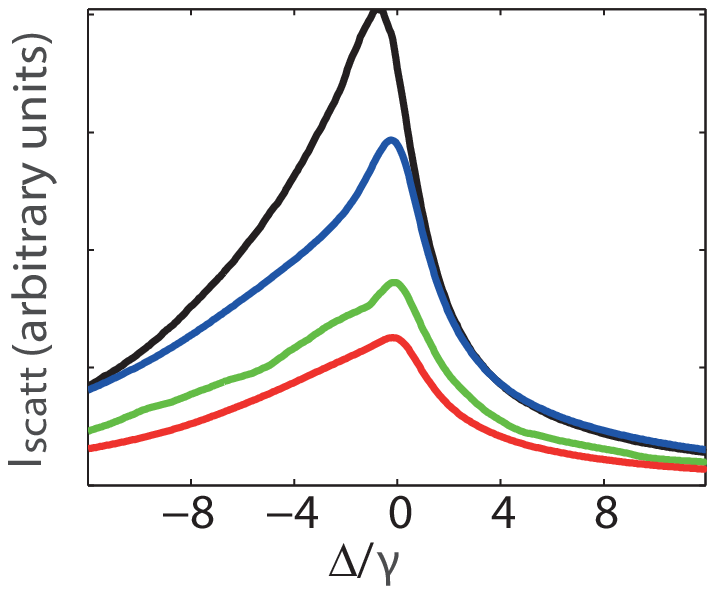}
    \includegraphics[width=0.48\columnwidth]{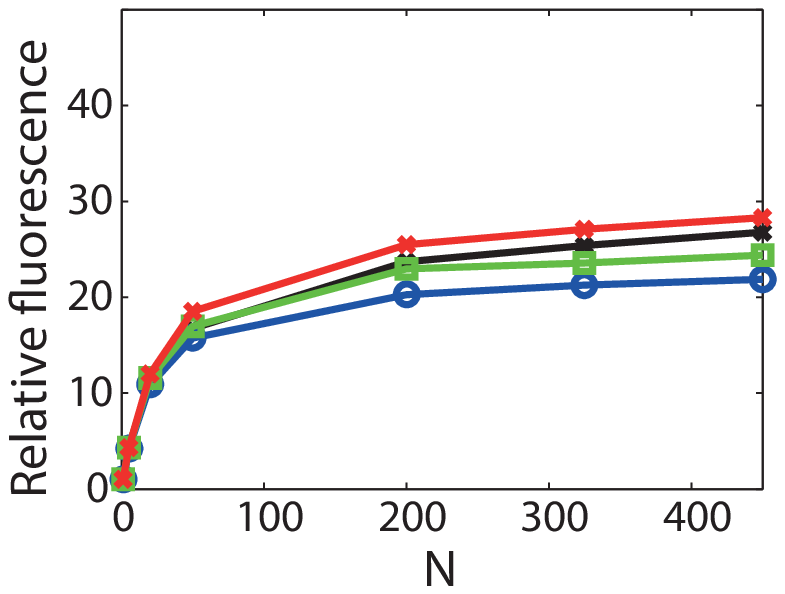}
  \caption{
  Left: The effects of Zeeman state populations on scattered
    intensity for an ensemble of $N=450$ $^{87}$Rb  atoms.
    The curves from top to bottom: the scattered intensity from two-level atoms (black), from $^{87}$Rb with
    $p_2=1$ (blue),     $p_2 = p_1 =p_0 =1/3$ (green), and $p_j =1/5$ for all $j$ (red).
    Right: Relative fluorescence as a function of number of atoms, for
    an incident field tuned to a single atom's peak fluorescence.
    The scattered intensity from an
    ensemble of atoms is normalized to the maximum of the
    scattered intensity from a single atom.  The curves from top to bottom:
    the scattered intensity from $^{87}$Rb with $p_j =1/5$ for all $j$ (red), two-level atoms (black), $^{87}$Rb with $p_2 = p_1 =p_0 =1/3$ (green),
    and $p_2=1$ (blue). The magnetic field is equal to zero.}
  \label{fig:I_scatt_pol_comp_B0}
\end{figure}

Figure~\ref{fig:I_scatt_pol_comp_B0}  shows both the fluorescence line shape for a fixed large ($N=450$) number of atoms and the fluorescence intensity on resonance for a variable number of atoms for three different initial level populations of Rb, and indeed for the hypothetical two-level atom. The left-hand side panel shows that the level structure has a significant quantitative effect on the fluorescence intensity, but not on the line shift that was the object of our comparisons with the experiments. In the right-hand side panel the fluorescence intensities for different atom numbers are normalized to the maximum fluorescence intensity for a single atom with the same level structure and initial state. There is an effect from the level structure, but, when viewed in this way, it is not particularly dramatic.

In addition, we have tested the robustness of the simulation results to various other parameters in the experiments and have found no notable changes in the resonance
shifts. For example, the Zeeman splitting due to the 1G magnetic field in the simulations does not significantly modify the response as compared with the zero field case.
In a dense ensemble of $^{87}$Rb atoms the main effect is to provide a slightly narrower resonance peak with a more recognizable `hump' on the red-detuned side of the peak
than in the absence of the magnetic field.

\section{Concluding remarks}
\label{sec:conclusion}

It has previously been shown in the case of coherent light transmission that
in dense, homogeneously-broadened atomic ensembles strong light-mediated interactions and the resulting light-induced
correlations can lead to a dramatic and qualitative failure of standard continuous effective-medium electrodynamics~\cite{Javanainen2014a,JavanainenMFT}.
This is because the standard textbook theory of optics~\cite{Jackson,BOR99} represents an effective-medium mean-field theory that
assumes each atom interacting with the average behavior of the surrounding atoms. In such models the spatial information about the
precise locations of the pointlike atoms -- and the corresponding details of the position-dependent DD interactions -- is washed out,
resulting in the absence of the light-induced correlations.

In the presence of inhomogeneous broadening the results of the standard electrodynamics of continuous polarizable medium, however, can be restored~\cite{Javanainen2014a}.
At sufficiently high temperatures, the simulations qualitatively agree with the standard established models of resonance line shifts, the LL shift and its similarly
mean-field theoretical collective (finite-size) counterpart, the ``cooperative Lamb shift''~\cite{Friedberg1973}.
The microscopic mechanism for the emergence of the continuous medium electrodynamics is the suppression of the
light-mediated resonant DD interactions between the atoms: With increasing inhomogeneous broadening the atoms are simply farther away from resonance
with the light scattered by the other atoms.
Formally, one can show~\cite{Javanainen2014a} that in thermal ensembles each recurrent scattering event is suppressed by the factor $\sim \gamma/\Omega$, where $\Omega$ denotes the
width of the inhomogeneous broadening.

Here we have explained in detail and extended our work of Refs.~\cite{Pellegrino2014a,Jenkins_thermshift} on near-resonance light scattering from small clouds of cold or thermal atoms.
We illustrated a substantially different behavior of resonance fluorescence of trapped, cold Rb atoms from that of thermal atoms.
In our analysis we performed side-by-side comparisons between experimental observations and large-scale
numerical simulations of resonance fluorescence in a dense cloud of $^{87}$Rb atoms.
Both the experiment and stochastic simulations demonstrate the emergence of collective DD interactions
that dramatically alter the optical response as the number of atoms is gradually increased.
We found that both the cold-atom simulations and the experimental observations of the
resonance line shifts and the total collected scattered light intensity substantially deviate from those
of thermal atomic ensembles. In particular, a density-dependent resonance shift of a thermal
atomic ensemble is almost entirely absent in a cold atomic cloud.
Our numerical models of fluorescence are also more involved than those of light transmission in Ref.~\cite{Javanainen2014a}, and incorporate the experimental setup in detail, including the inhomogeneous atom densities
due to the trapping potential, the internal multilevel structure of $^{87}$Rb, and the imaging geometry with the optical components
(e.g., lenses and polarizers).

Moreover, we analyzed the effect of strong light-mediated interactions between the atoms by calculating
the collective radiative excitation eigenmodes of the system.
As a result of light-induced DD interactions, the response of the sample becomes collective, exhibiting collective
radiative resonance linewidths and line shifts including those with
subradiant and superradiant character. When the collective radiative decay rates are far from those
of a single isolated atom, the optical response of the cloud cannot be approximated by the one consisting of independent atoms,
and a broad distribution of decay rates is an indication of strong DD interactions.

The role of collective excitation eigenmodes in the optical response is in particular illustrated by the
calculation of the temporal profile of the decay of light-induced excitations after the incident laser pulse is switched off.
At high atom densities the simulations predict a significantly slower decay
for cold than for thermal atom clouds. In a logarithmic scale the calculated scattered power notably deviates from a straight line,
indicating non-negligible occupations of (both superradiant and subradiant) collective modes with different
radiative linewidths.

\acknowledgments

We acknowledge discussions with M.\ D.\ Lee. This work was financially supported by the Leverhulme Trust, EPSRC, NSF Grants
No.\ PHY-0967644 and No.\ PHY-1401151, and the EU through the ERC Starting Grant ARENA
and the HAIRS project, the Triangle de la Physique (COLISCINA project), the labex PALM (ECONOMIC project),
and the Region Ile-de-France (LISCOLEM project).

\end{document}